\documentstyle[12pt,epsf]{article}
\newcommand{\aem}{\alpha}
\newcommand{\RE}{{\rm Re}}
\newcommand{\IM}{{\rm Im}}
\newcommand{\mc}{m_{\rm c}}
\newcommand{\mt}{m_{\rm t}}
\newcommand{\dS}{\Delta S=1}
\newcommand{\as}{\alpha_{\rm s}}
\newcommand{\mvs}{\vbox{\vskip 8mm}}
\newcommand{\Heff}{{\cal H}_{\rm eff}}
\newcommand{\Kpipi}{K \rightarrow \pi\pi}
\newcommand{\epe}{\varepsilon'/\varepsilon}
\newcommand{\Lms}{\Lambda_{\overline{\rm MS}}}
\newcommand{\lsim}{~{}_{\textstyle\sim}^{\textstyle <}~}
\newcommand{\gsim}{~{}_{\textstyle\sim}^{\textstyle >}~}
\begin{document}
\
\thispagestyle{empty}
\vskip 2cm
\begin{flushright}
June 1993
\end{flushright}
\vskip 2cm
\begin{center}
{\large \bf THE ANATOMY OF $\epe$ BEYOND LEADING LOGARITHMS
 WITH IMPROVED HADRONIC MATRIX ELEMENTS\footnote{Supported by the
 German Bundesministerium f\"ur Forschung und Technologie under
 contract 06 TM 732 and by the CEC Science project SC1-CT91-0729.}}
\end{center}
\bigskip
\begin{center}
{ {\it Matthias Jamin} \\ 
Theory Division, CERN, CH-1211 Gen\` eve 23, Switzerland}
\end{center}
\vskip 2.5cm
\noindent {\bf Abstract}
\medskip

\noindent
Results of a recent calculation of the effective $\dS$ Hamiltonian
at the next-to-leading order will be presented\footnote{Work done in
collaboration with A. J. Buras and M. E. Lautenbacher.}. These, together
with an improved treatment of hadronic matrix elements, are used to
evaluate the measure for direct CP-violation, $\epe$, at the
next-to-leading order. Taking $\mt=130\,GeV$, $\Lms=300\,MeV$ and
calculating $\langle Q_6 \rangle$ and $\langle Q_8 \rangle$ in the
$1/N$ approach, we find in the NDR scheme $\epe = (6.7 \pm 2.6)\times 10^{-4}$
in agreement with the experimental findings of E731.  We point out however
that the increase of $\langle Q_6 \rangle$ by only a factor of two gives
$\epe = (20.0 \pm 6.5)\times 10^{-4}$ in agreement with the result of NA31.
The dependencies of $\epe$ on $\Lms$, $\mt$, and some $B$-parameters,
parameterizing hadronic matrix elements, are briefly discussed.
\vskip 1.5cm
\begin{center}
{\it Invited talk given at the XXVIII Rencontres de Moriond, Les Arcs, 1993}
\end{center}
\newpage \noindent
\setcounter{page}{1}
\noindent {\bf 1 Introduction}
\medskip

In recent years, determinations of the measure for direct CP-violation
in $\Kpipi$ decays, $\epe$, both on the experimental and on the
theoretical side, have experienced great improvement, though the
situation, nevertheless, is not yet conclusive. The experimental
researchers, after heroic efforts on both sides of the atlantic,
find \cite{barrwinstein:91,gibbons:93},
\begin{equation}
\RE(\epe) \; = \; \left\{
\begin{array}{ll}
(23 \pm 7)    \cdot 10^{-4} & \hbox{NA31} \\
(7.4 \pm 6.0) \cdot 10^{-4} & \hbox{E731}
\end{array} \right. \, ,
\label{eq:1}
\end{equation}
clearly indicating a non-zero $\epe$ in the case of NA31, whereas the
value of E731 is still compatible with superweak theories in which
$\epe=0$.
 
Theoretically, $\epe$ is governed by penguin contributions with QCD
penguins dominating for values of $\mt\lsim 150\,GeV$, but if the top
quark mass would turn out to be as large as ${\cal O}(200\,GeV)$, as
has been first pointed out be Flynn and Randall \cite{flynn:89},
QED penguins become important and tend to cancel the QCD contribution,
yielding $\epe$ close to zero.

At the leading order a detailed anatomy of $\epe$ in the presence of a heavy
top quark has been performed by the authors of ref.~\cite{buchallaetal:90},
which has subsequently been corroborated in \cite{paschos:91} and
\cite{lusignoli:92}. Since the outcome of the fight between QCD and
electroweak penguins is rather sensitive to the various approximations
used in \cite{buchallaetal:90,paschos:91,lusignoli:92}, it is very
important to improve the theoretical calculations both on the
short-distance side (Wilson coefficient functions $C_i(\mu)$) and
on the long distance side (hadronic matrix elements
$\langle Q_i(\mu) \rangle$). These improvements will be outlined in
the following.

\bigskip

\noindent {\bf 2 Coefficient Functions}
\medskip
 
In the framework of the operator product expansion, $\dS$ weak transitions
are described by the effective Hamiltonian
\begin{equation}
\Heff(\dS) \; = \; \frac{G_F}{\sqrt{2}}\,\sum_{i=1}^{10}\, C_i(\mu)
\,Q_i(\mu) \,,
\label{eq:2}
\end{equation}
with $C_i(\mu)$ being the Wilson coefficient functions. The operators
$Q_i$ can be classified into $Q_{1,2}$ being current-current, $Q_{3-6}$
QCD penguin, and $Q_{7-10}$ electroweak penguin operators
\cite{buchallaetal:90}.

The next-to-leading order calculation of the coefficient functions
$C_i(\mu)$ now consists of the following steps:
\begin{itemize}
\item calculation of the $10\times10$ ${\cal O}(\as^2)$ and ${\cal O}(\aem\as)$
 anomalous dimension matrices which govern the next-to-leading order
 renormalization group evolution from some high energy scale ${\cal O}(M_W)$
 down to the low energy scale $\mu$ \cite{burasetal:92b,burasetal:92c},
\item calculation of the initial conditions $C_i(M_W)$ which are used
 as a starting point for the renormalization group evolution
 \cite{burasetal:92a,burasetal:93},
\item and solving the renormalization group equation with inclusion
 of both ${\cal O}(\as^2)$ and ${\cal O}(\aem\as)$ corrections
 \cite{burasetal:92a,burasetal:93}.
\end{itemize}
\noindent
An independent calculation of the ${\cal O}(\as^2)$ and ${\cal O}(\aem\as)$
anomalous dimension matrices has also been performed by the authors of
\cite{martinellietal:93}, finding agreement with our final results.

Writing $C_i(\mu)$ as $\lambda_{\rm u}\,z_i(\mu)-\lambda_{\rm t}\,y_i(\mu)$,
with $\lambda_{\rm i}=V_{id}V_{is}^{*}$, $V_{ij}$ being
Cabibbo-Kobayashi-Maskawa (CKM) matrix elements, splits $C_i(\mu)$
into the major real part $\sim z_i(\mu)$,
describing CP-conserving transitions, and the imaginary part
$\sim y_i(\mu)$, being responsible for CP-violating transitions.
$z_i$ and $y_i$ depend on $\Lms$, the renormalization scale $\mu$, the
renormalization scheme, and in the case of $y_i$ also on $\mt$. A
thorough discussion of all these dependencies has been presented in
ref. \cite{burasetal:93}. As an example, we explicitly give in
tab.~\ref{tab:1} the coefficient functions for $\Lms=300\,GeV$,
$\mu=1\,GeV$, a renormalization scheme with anticommuting $\gamma_5$
(NDR), and $\mt=130\,GeV$.
\begin{table}[thb]
\begin{center}
\begin{tabular}{|c|c|c|c|c|c|c|c|c|c|}
\hline
$z_1$ & $z_2$ & $z_3$ & $z_4$ & $z_5$ & $z_6$ &
$z_7/\aem$ & $z_8/\aem$ & $z_9/\aem$ & $z_{10}/\aem$ \\
\hline
-0.486 & 1.262 & 0.013 & -0.035 & 0.007 & -0.035 &
 0.008 & 0.016 & 0.016 & -0.009 \\
\hline
\hline
$y_1$ & $y_2$ & $y_3$ & $y_4$ & $y_5$ & $y_6$ &
$y_7/\aem$ & $y_8/\aem$ & $y_9/\aem$ & $y_{10}/\aem$ \\
\hline
0 & 0 & 0.029 & -0.052 & 0.001 & -0.099 & -0.080 & 0.095 & -1.225 & 0.502 \\
\hline
\end{tabular}
\end{center}
\caption{The coefficient functions $z_i(1\,GeV)$ and $y_i(1\,GeV)$.
\label{tab:1}}
\end{table}

\bigskip
 
\noindent {\bf 3 Hadronic Matrix Elements}
\medskip

Besides the coefficient functions $y_i(\mu)$, for the calculation
of $\epe$ we also need the matrix elements $\langle\pi\pi|Q_i(\mu)|K
\rangle_{0,2}$, where the subscript denotes the isospin of the final
state pions. Since a direct calculation of the matrix elements involves
long-distance dynamics in QCD, and is therefore rather difficult and
uncertain, in \cite{burasetal:93} we advocated a more phenomenological
approach. For further references on other non-perturbative approaches
to hadronic matrix elements see also \cite{burasetal:93}.

Imposing experimental data on CP-conserving $\Kpipi$ decays and some
plausible properties of the hadronic matrix elements which are
fulfilled by all common non-perturbative methods, we can fix part
of the matrix elements completely and express the rest in terms of
the three parameters $B_2^{(1/2)}(\mu)$, $B_6^{(1/2)}(\mu)$, and
$B_8^{(3/2)}(\mu)$. In addition, from experimental data we can deduce
$B_2^{(1/2)}(\mc)=6.7\pm0.9$ in the NDR scheme. In our approach it
is most convenient to evaluate the matrix elements at the scale $\mc$.

\bigskip
 
\noindent {\bf 4 Main Results for $\epe$}
\medskip

Using the results for the coefficient functions $y_i$ and the hadronic
matrix elements from sects.~2 and 3 respectively \cite{burasetal:93},
we can write the CP-violating quantity $\epe$ in the following form,
\begin{equation}
\frac{\varepsilon'}{\varepsilon} \; = \; 10^{-4}\,\left[\frac{\IM
\lambda_{\rm t}}{1.7\cdot 10^{-4}}\right]\, \left[\,P^{(1/2)}-P^{(3/2)}
\,\right] \,.
\label{eq:3}
\end{equation}
We have factored out the central value of $\IM\lambda_{\rm t}$ in the
range for the CKM-matrix elements considered in \cite{burasetal:93},
such that $\epe$ (in units of $10^{-4}$) is directly given by $P^{(1/2)}$
and $P^{(3/2)}$. The contribution from $\Delta I=1/2$, $P^{(1/2)}$, is
dominated by QCD penguins, whereas the $\Delta I=3/2$ part mainly comes
from electroweak penguins. In terms of the $B$-parameters introduced in
the last section, their expressions are:
\begin{eqnarray}
P^{(1/2)} & = & a_0^{(1/2)} + a_2^{(1/2)}\,B_2^{(1/2)} +
                  a_6^{(1/2)}\,B_6^{(1/2)} \, ,
\label{eq:4} \\
\mvs
P^{(3/2)} & = & a_0^{(3/2)} + a_8^{(3/2)}\,B_8^{(3/2)} \, .
\label{eq:5}
\end{eqnarray}
The coefficients $a_i$ depend on $\Lms$, the renormalization scheme
considered, and the top quark mass. Again, as an example, in tab.~\ref{tab:2}
we give the values of the $a_i$ for $\Lms=300\,MeV$, the NDR scheme,
and $\mt=130$, $150$, and $170\,GeV$.
\begin{table}[thb]
\begin{center}
\begin{tabular}{|c|c|c|c||c|c|}
\hline
$\mt [GeV]$ & $a_0^{(1/2)}$ & $a_2^{(1/2)}$ & $a_6^{(1/2)}$ &
$a_0^{(3/2)}$ & $a_8^{(3/2)}$ \\
\hline
\hline
130 & -7.61 & 0.54 & 11.68 & -1.21 & 3.02 \\
\hline
150 & -7.24 & 0.51 & 11.77 & -1.40 & 4.59 \\
\hline
170 & -6.84 & 0.47 & 11.85 & -1.59 & 6.38 \\
\hline
\end{tabular}
\end{center}
\caption{Coefficients in the expansion of $P^{(1/2)}$ and $P^{(3/2)}$.
\label{tab:2}}
\end{table}

Using these coefficients together with the factorization values for
$B_6$ and $B_8$, $B_6^{(1/2)}(\mc)=B_8^{(3/2)}(\mc)=1$, and $\mt=130\,GeV$,
our final results for $\epe$ are
\begin{equation}
\epe = \left\{
\begin{array}{ll}
(6.7 \pm 2.6) \times 10^{-4} & \qquad 0 < \delta \le \frac{\pi}{2} \, , \\
(4.8 \pm 2.2) \times 10^{-4} & \qquad \frac{\pi}{2} < \delta < \pi  \, ,
\end{array}
\right.
\label{eq:6}
\end{equation}
in perfect agreement with the E731 result. However, for
$B_6^{(1/2)}(\mc)=2$ and $B_8^{(3/2)}(\mc)=1$ we find
\begin{equation}
\epe = \left\{
\begin{array}{ll}
(20.0 \pm 6.5) \times 10^{-4} & \qquad 0 < \delta \le \frac{\pi}{2} \, ,\\
(14.4 \pm 5.6) \times 10^{-4} & \qquad \frac{\pi}{2} < \delta < \pi \, ,
\end{array}
\right.
\label{eq:7}
\end{equation}
in agreement with the findings of NA31. Here, $\delta$ is the complex
phase in the CKM matrix, and the explicit errors include variation of
the CKM-matrix elements and $B_2^{(1/2)}$ as given in \cite{burasetal:93}
and sect.~3 respectively.

\begin{figure}[hbt]
\centerline{
\epsfysize=2.3in
\epsffile{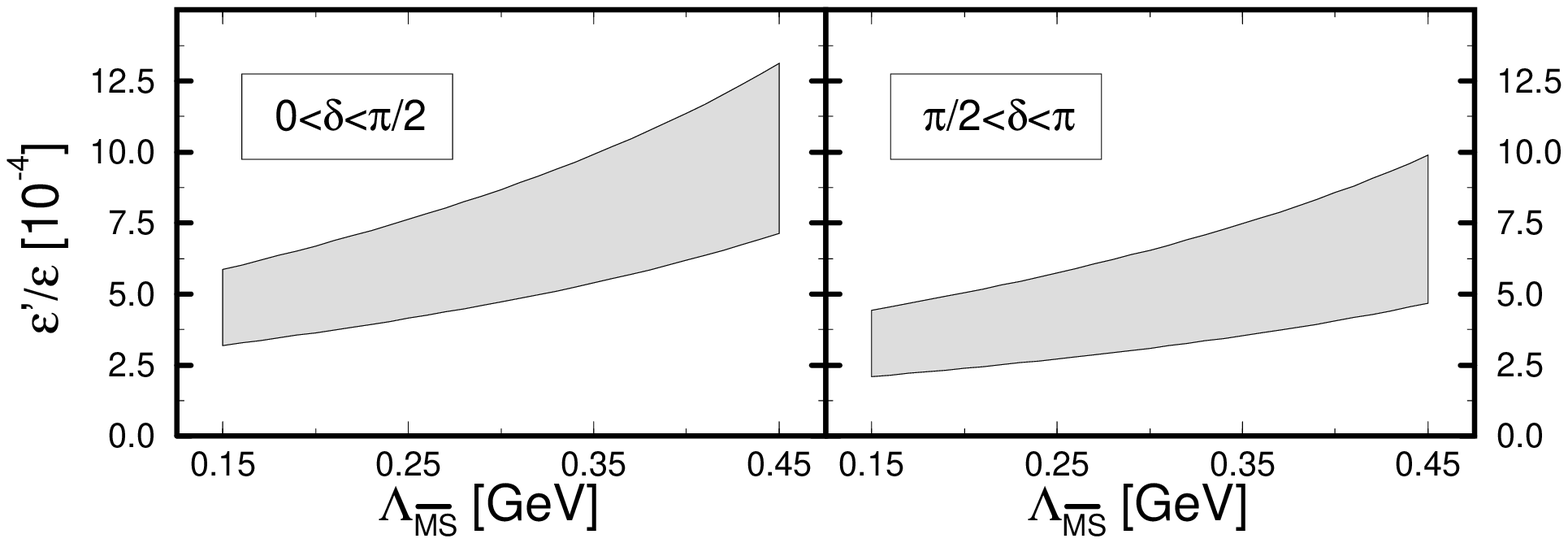}
\vspace{0.1in}
}
\caption[]{$\epe$ as a function of $\Lms$ for $\mt=130\,GeV$ and $B_6=B_8=1$.
\label{fig:1}}
\end{figure}

Let us briefly point out further dependencies of $\epe$. For a thorough
discussion the reader is again referred to ref.~\cite{burasetal:93}.
\begin{itemize}
\item $\epe$ decreases if $\mt$ increases but becomes zero only
 for $\mt\gsim 200\,GeV$.
\item $\epe$ increases if $\Lms$ increases. This dependence is
 illustrated in fig.~\ref{fig:1} for $\mt=130\,GeV$ and $B_6=B_8=1$.
\end{itemize}
Similar final results for $\epe$ were also obtained in ref.
\cite{martinellietal:93a}. Taken separately however, some
contributions to $\epe$ differ notably from the values obtained in
our analysis \cite{burasetal:93}.
\bigskip
 
\noindent {\bf 5 Conclusions}
\medskip

Despite the achievements in recent years, the fate of theoretical
estimates of $\epe$ in the years to come depends crucially on whether
it will be possible to further reduce the uncertainties in $\mt$,
$\Lms$, $B_6^{(1/2)}$, $B_8^{(3/2)}$, and the CKM-matrix elements.

On the other hand, with the new generation of measurements, in a few
years of time hopefully also the experimental situation will be clarified.
Nevertheless, if the top quark turns out to be too heavy, this may
become difficult.

\newpage

\noindent{\bf Acknowledgements}
\medskip

The author wishes to thank Andrzej Buras and Markus Lautenbacher
for most enjoyable collaboration.

\end{document}